\documentclass[pra,showpacs]{revtex4}% Physical Review A

\usepackage{graphicx}% Include figure files
\usepackage{dcolumn}% Align table columns on decimal point
\usepackage{bm}% bold math

\begin{document}

\title{On a Density-of-States Approach to Bohmian Mechanics}% Force line breaks with \\

\author{Guy Potvin}
\email{guy.potvin@drdc-rddc.gc.ca}
\affiliation{Defence R\&D Canada - Valcartier, Val-B\'{e}lair, Qu\'{e}bec, Canada, G3J 1X5}

\date{\today}% It is always \today, today,
             %  but any date may be explicitly specified

\begin{abstract}
We propose the idea that in Bohmian mechanics the wavefunction is related to a density of states and explore some of its consequences.
Specifically, it allows a maximum-entropy interpretation of quantum probabilities, which creates a stronger link between it and statistical mechanics.
The proposed approach also allows a range of extensions of the guidance condition in Bohmian mechanics.
\end{abstract}



\maketitle

\section{Introduction}

Bohmian mechanics is a determistic pilot-wave theory of quantum mechanics \cite{Bohm52a,Bohm52b,Holland93}.
As such, it has a dual ontology: there exists an actual $3N$-dimensional configuration of $N$ material particles, ${\bf \tilde{x}}$, moving in a $3N$-dimensional physical configuration space, ${\bf x}$, due to the deterministic guidance of a wavefunction defined over the configuration space, $\psi({\bf x})$, also considered a real object.
For non-relativistic spinless particles, the wavefunction obeys the Schroedinger equation,
\begin{equation}
i \hbar \frac{\partial \psi}{\partial t} = -\frac{\hbar^{2}}{2m_{k}}{\bf \nabla}^{2}_{k} \psi + V({\bf x}) \psi
\label{E_Schr}
\end{equation}
where the repeated index $k$ denotes a summation over the $N$ particles and $V$ is an arbitrary potential.
Equation (\ref{E_Schr}) can be converted into a continuity equation,
\begin{equation}
\frac{\partial}{\partial t} |\psi|^{2} + {\bf \nabla} \cdot {\bf F} = 0
\label{E_Cont}
\end{equation}
where ${\bf F}$ is the wavefunction flux in configuration space.
\begin{equation}
{\bf F} = \frac{\hbar}{2im_{k}}(\psi^{*} {\bf \nabla}_{k} \psi - \psi {\bf \nabla}_{k} \psi^{*})
\label{E_Flux}
\end{equation}
The velocity of the particles in configuration space, ${\bf \tilde{v}} = {\rm d}{\bf \tilde{x}}/{\rm d} t$ is given by the  condition
\begin{equation}
{\bf \tilde{v}} = \frac{{\bf F}({\bf \tilde{x}})}{|\psi({\bf \tilde{x}})|^{2}}.
\label{E_Guide}
\end{equation}
In other words, the particles follow the streamlines of the wavefunction.\\

These ontological statements are supplemented with a probabilistic one, namely that the particle configuration is randomly distributed in the configuration space such that $\rho({\bf \tilde{x}})=|\psi({\bf x})|^{2}$ (i.e. Born's rule or quantum equilibrium (QE) \cite{Berndl.et.al95}).
The central idea of this paper is that the number of possible particle configurations, $\tilde{n}$, in a physical configuration volume $V$ is not proportional to that volume, but rather
\begin{equation}
\tilde{n} \propto \int_{V} |\psi({\bf x})|^{2} {\rm d}V .
\label{E_idea}
\end{equation}
In other words, let us suppose the existence of a discrete lattice of possible configuration states for the particles.
This lattice is unevenly spaced such that some regions of configuration space will contain more lattice points than others, such as in Eq. (\ref{E_idea}).
We then take the limit of an infinitely fine lattice such that any given region of configuration space contains an infinite number of points, but we can normalize this number so as to create a continuous density function.\\

Finally, we introduce a $3N$-dimensional quantum configuration space ${\bf q}$ in which the number of possible particle configurations, ${\bf \tilde{q}}$, in an arbitrary volume $V_{q}$ is always proportional to that volume.
The quantum space (or q-space) is therefore `uniform' in the sense that in it our hypothetical lattice is evenly spaced.
If we further suppose the existence of a coordinate transformation linking the quantum and real spaces, ${\bf q} = {\bf q}({\bf x})$ and ${\bf x} = {\bf x}({\bf q})$, we can say that a density-of-states function $\omega$ is equal to the Jacobian of the transformation,
\begin{equation}
\omega = \frac{{\rm d}V_{q}}{{\rm d}V} = J\left(\frac{{\bf q}}{{\bf x}}\right).
\label{E_Jac}
\end{equation}
We propose the identity
\begin{equation}
|\psi({\bf x})|^{2} = C \omega({\bf x})
\label{E_link}
\end{equation}
where $C$ is a normalizing constant.
We will explore some of the consequences of this view in the following sections.
We begin with statistical mechanical considerations in Sec. \ref{S_ME}, followed by kinematic considerations in Sec. \ref{S_MiQS}.
We end with a discussion about this aproach in Sec. \ref{S_Disc}.

\section{Maximum Entropy}
\label{S_ME}

Entropy, in the information-theoretic sense formulated by Shannon \cite{Shannon48}, expresses our ignorance about a physical situation, or equivalently, the amount of information that is obtained when we ascertain that situation.
For a discrete random variable, ($X = [x_{1},x_{2},...,x_{n}]$, where $x_{1}$ occurs with probability $p_{1}$, $x_{2}$ with $p_{2}$ and so on) the entropy of its distribution is
\begin{equation}
S = -\sum_{i=1}^{n}p_{i} \ln p_{i}.
\label{E_Ent}
\end{equation}
If we also happen to know that the distribution must satisfy some constraints, $f_{k} = \sum p_{i} x_{i}^{k}$, then the distribution with the highest entropy that satisfies these constraints is the most unbiased inference we can make about the physical situation given our information \cite{Jaynes03}.\\

There is also an important combinatorial argument in favor of maximum entropy distributions.
If we have $M$ independent observations of $X$, such that we have a sequence of random numbers $d = [X_{1},...,X_{M}]$.
For each such sequence we can create a histogram $h = [m_{1},...,m_{n}]$, where $m_{i}$ is the number of times the value $x_{i}$ was obtained in the sequence (and therefore $\sum m_{i} = M$, we also assume that the histograms respect the constraints imposed on the probability distribution).
For each histogram we can attribute an entropy $S_{h} = -M^{-1} \sum m_{i} \ln (m_{i}/M)$.
It turns out that for large $M$, the number of possible sequences $d$ that result in a given histogram $h$ is \cite{Jaynes68}
\begin{equation}
W \propto \exp [M S_{h}].
\label{E_Combo}
\end{equation}
This means that the histogram with the highest entropy occupies the greatest `volume' in the space of possible data sequences.
Conversely, those histograms with a smaller entropy occupy a volume that is exponentially smaller.
It is important to note that the entropy functional $S$ in Eq. (\ref{E_Ent}) refers to `subjective' probabilities expressing our beliefs about $X$ whereas $S_{h}$ refers to observed relative frequencies.
These are two distinct but complementary concepts (see Jaynes \cite{Jaynes03} for a discussion on the relationship between them and see Sklar \cite{Sklar93} for a discussion on the different views on probability in physics).\\

For a continuous random variable, the entropy is not as straightforward as for the discrete variable.
This is because Shannon's original work dealt with the capacity of a channel to communicate a message consisting of a discrete sequence of letters, each a member of a discrete alphabet.
In order to make progress, Shannon had to completely eliminate the semantic aspect of the message and focus on the purely technical problem of accurately transmitting an arbitrary message.
Since Shannon removed all considerations of not only the content of the message, but also what langage it is in or even whether the message is encrypted, he had to consider each member of the alphabet as carrying the same amount of information, or being \emph{a priori} equally likely.
In order to pass to the continuous case, we must consider a discrete set of values which the random variable may take, $X = [x_{1},x_{2},...,x_{n}]$.
These possible values are not necessarily evenly spaced.
Then, taking the limit as before of an infinite number of points in an infinitely fine lattice, we obtain the expression for the entropy of a continuous variable, \cite{Jaynes03,Jaynes68,Jaynes63}
\begin{equation}
S = -\int \rho(x) \ln \left[\frac{\rho(x)}{m(x)}\right] {\rm d} x
\label{E_Ent_c}
\end{equation}
where $m(x)$ is the continuous density of states that are \emph{a priori} equally likely.
It also acts as a measure which makes the entropy invariant with respect to coordinate transformations.\\

If we also impose $K$ constraints $f_{k} = \int \rho(x) x^{k} {\rm d}x$, then the maximum entropy density is
\begin{equation}
\rho(x) = \frac{m(x)}{Z} \exp\left[-\sum_{k=1}^{K}\lambda_{k} x^{k}\right]
\label{E_MEcd}
\end{equation}
where $Z$ is the partition function that normalizes the density and the $\lambda_{k}$ are Lagrange multipliers that ensure the proper moments for the density.
It is clear from Eq. (\ref{E_MEcd}) that if no constraints are imposed, the density $m$ represents that state of complete ignorance about $x$ (or the application of Bernoulli's ``Principle of Insufficient Reason" or Keynes' ``Principle of Indifference" \cite{Jaynes79}).\\

As the reader may have guessed, the entropy for the Bohmian particle probability density, $\rho({\bf \tilde{x}})$, is given in terms of the density of states $\omega({\bf x})$:
\begin{equation}
S = -\int \rho({\bf \tilde{x}}) \ln \left[\frac{\rho({\bf \tilde{x}})}{\omega({\bf x})}\right] {\rm d} V .
\label{E_Ent_B}
\end{equation}
In fact, something like Eq. (\ref{E_Ent_B}) has been written before by Valentini \cite{Valentini91a,Valentini91b,Valentini04}, but in the sense of a cross-entropy between $\rho$ and $|\psi|^{2}$.
In other words, as a measure of how different two distributions are and not as a measure of information with respect to a set of \emph{a priori} equally likely states.
Valentini used a sub-quantum H-function, related to Eq. (\ref{E_Ent_B}), to demonstrate a sub-quantum H-theorem where QE arises in Bohmian mechanics as a result of dynamical mixing and coarse-graining arguments.\\

If we do not impose any further constraints, then the least biased probability density that takes into account the proper counting of states (i.e. maximum entropy) is $\rho({\bf \tilde{x}})=|\psi({\bf x})|^{2}$.
For what follows, it will be useful to make a few assumptions, namely that the `universe' under consideration is finite such that the total volume $V = L^{3N}$, where $L$ is very large and we neglect any surface effects.
We further assume that q-space possesses the same volume as real space $V_{q} = V$ such that,
\begin{equation}
\int_{V} \omega({\bf x}) {\rm d}V = V
\label{E_Omega}
\end{equation}
meaning that compressions $\omega > 1$ and rarefactions $\omega < 1$ cancel (and that the normalizing constant in Eq. (\ref{E_link}) is $C = V^{-1}$).
We now consider an idealized experiment where we have a large number of particles $N$, each one prepared in the state $\psi$.
For simplicity, we assume that the particles exist in one dimension $x$.
The experiment consists in measuring each particle's position to within a series of adjacent and non-overlapping bins of equal width $\delta x$, and then compiling the results into a histogram.
From the previous combinatorial argument, it is clear that those histograms that are close to QE (maximum entropy) occupy a very large proportion of q-space and those histograms that deviate significantly from QE occupy a very small proportion of q-space.\\

There is a direct analogy between our volume in q-space argument for quantum equilibrium and the notion of \emph{typicality} developed by D\"{u}rr, Goldstein and Zanghi (DGZ) \cite{Durr.et.al92}.
Let us suppose we have a set of initial particle configurations $A$, which we assume is piecewise smooth.
The typicality of this set is given as
\begin{equation}
T(A) = \int_{A} |\psi({\bf x})|^{2} {\rm d} V
\label{E_Typ}
\end{equation}
where $|\psi({\bf x})|^{2}$ acts as a measure of typicality.
Since we assume that the wavefunction is normalized, we can say that \emph{typical} sets are those whose typicality is close to one.
Since Bohmian mechanics is a deterministic theory, the initial conditions also denote a set of trajectories.
What DGZ have shown is that the configuration trajectories that give relative frequencies in repeated experiments (in space and time) close to quantum equilibrium form a set whose typicality is very close to one in the sense of Eq. (\ref{E_Typ}).
However, from Eqs. (\ref{E_Jac}) and (\ref{E_link}) it is clear that typicality is proportional to the volume in q-space of the set.
Therefore, the set of trajectories that produce relative frequencies close to QE occupy most of q-space, exactly the result obtained previously using the combinatorial argument.
We have therefore reformulated DGZ's result in geometrical terms.\\

There is also a direct analogy with Boltzmann's views on classical statistical mechanics \cite{Lebowitz93}.
Consider, for instance, an isolated ideal gas contained in a volume $V$ with a total energy $E$.
The microscopic state of the gas, namely the positions and momenta of all $N$ particles where $N$ is a very large number, is therefore constrained on a constant-energy surface in phase space.
The macroscopic state of the gas, for instance the density field of the gas within the container, corresponds to a certain volume in phase space.
It can be shown that macroscopic states corresponding to equilibrium (i.e uniform density) occupy a much greater volume in phase space than those associated with non-equilibrium (non-uniform density).
Given its much greater volume, it is reasonable to expect that the microsopic state of the gas is in the equilibrium state, or, if the gas is prepared in a non-equilibrium state, that it will probably reach the equilibrium state and stay there for a very long time.
Similarly, if we consider the density of measured particle positions in an experiment as our macroscopic state, then it is reasonable to expect that the particle configuration in q-space, the microscopic state, will be in the very large volume corresponding to the QE state.\\

Statistical mechanics is, in large part, about counting microscopic states, but only as long as we count the right microscopic states.
For instance, Gibb's paradox \cite{Reif65} arises if we count each and every possible microscopic state, regardless of their indistinguishability.
The paradox disappears if we count indistinguishable states only once.
Furthermore, it is the quantum mechanical symmetry properties of the particles that tell us whether to use Fermi-Dirac or Bose-Einstein statistics \cite{Reif65}.
In other words, quantum mechanics determines the actual number of microscopic states available to perform statistical mechanical calculations.
Similarly, we propose that it is the configurations in q-space that are the proper microscopic states and quantum mechanics is about how they relate to the physical configuration space.
Indeed, the use of the term `density-of-states' is borrowed from statistical mechanics and is meant to emphasize the link between it and Bohmian mechanics.
In statistical mechanics, the density-of-states refers to the number of energy eigenstates within a given energy interval \cite{Reif65}.
However, there is nothing special about energy and energy eigenstates; they just happen to be relevant to the thermodynamical problem at hand.
In our case, it is the configuration states that are relevant and so we extend the density-of-states concept in that direction.

\section{Motion in Q-space}
\label{S_MiQS}

Bohmian mechanics is based on Madelung's hydrodynamic representation of quantum mechanics \cite{Holland93}.
This representation rewrites Schroedinger's equation into a continuity equation (\ref{E_Cont}) and a momentum conservation equation.
Equation (\ref{E_Cont}) is written in the spatial, or Eulerian, representation.
However, a fluid is made up of fluid particles and it is possible to write the hydrodynamic equations in a particle, or Lagrangian, representation \cite{Lamb93}.
This is done by postulating a Lagrangian coordinate system ${\bf a}$ that follows each fluid particle and so acts as a label for each particle.
The Lagrangian coordinate is related to the physical coordinate ${\bf x}$ by a time-dependent coordinate transformation ${\bf x} = {\bf x}({\bf a},t)$, such that for a fixed ${\bf a}$ the transformation describes the trajectory of that fluid particle.\\

So far, the concept of q-space has simply geometrized previously known results.
We now need to say how the q-space coordinates change with time, ${\bf x} = {\bf x}({\bf q},t)$.
To do this, we postulate that q-space is a Lagrangian variable that therefore follow the streamlines of the hydrodynamic representation,
\begin{equation}
\frac{\partial {\bf x}}{\partial t} \bigg{|}_{{\bf q}} = \frac{{\bf F}({\bf x}({\bf q},t))}{|\psi({\bf x}({\bf q},t))|^{2}}.
\label{E_Qguide}
\end{equation}
Comparing Eq. (\ref{E_Guide}) with Eq. (\ref{E_Qguide}), it is clear that in Bohmian mechanics the particles do not move in q-space.
This need not be the case.
Indeed, in hydrodynamics neighboring fluid particles constantly exchange molecules, so that although the fluid particles do not move in the Lagrangian coordinate system, a given molecule would perform a random walk in it.
Similarly, we can suppose that the particles undergo a random movement in q-space in such a way that a uniform probability density in q-space is stationary.
Such a theory would be empirically equivalent to Bohmian mechanics because if the particles are initially uniformly distributed over q-space (which corresponds to QE), such a movement would not change such a distribution.
By `random movement' we could mean a stochastic process, but we could also mean that the particles move with a constant velocity in q-space that is randomly distributed uniformly over all directions.
Whatever its nature, we will not elaborate further on this movement and will consider only the conditional probability density that the particles have a q-space configuration ${\bf {\tilde q}}_{2}$ at a later time $t_{2}$ given that they were at the configuration ${\bf {\tilde q}}_{1}$ at an earlier time $t_{1}$, $\rho_{q} ({\bf {\tilde q}}_{2},t_{2}|{\bf {\tilde q}}_{1},t_{1})$.
A fairly general and reasonable condition to ensure a stationary uniform density is the translation invariance of the condition probability density (again we ignore surface effects).
\begin{equation}
\rho_{q} ({\bf {\tilde q}}_{2},t_{2}|{\bf {\tilde q}}_{1},t_{1}) = f({\bf {\tilde q}}_{2}-{\bf {\tilde q}}_{1},t_{1},t_{2})
\label{E_Transinv}
\end{equation}
We can impose the additional constraint that there be no average displacement,
\begin{equation}
\int {\bf {\tilde q}}_{2} \rho_{q} ({\bf {\tilde q}}_{2},t_{2}|{\bf {\tilde q}}_{1},t_{1}) {\rm d} {\bf {\tilde q}}_{2} = {\bf {\tilde q}}_{1}
\label{E_Noflux}
\end{equation}
for any $t_{2}$ and $t_{1}$.
This constraint is desirable because we recover Bohmian mechanics on average.
Another desirable constraint would be to make the rate of increase of the displacement variance inversely proportional to some power of the mass, since this would give stability to macroscopic objects.\\

The corresponding conditional probability density for the physical configuration is
\begin{equation}
\rho ({\bf {\tilde x}}_{2}, t_{2} | {\bf {\tilde x}}_{1}, t_{1}) = J\left(\frac{{\bf q}}{{\bf x}}\right) \rho_{q} ({\bf q}({\bf {\tilde x}}_{2}),t_{2}|{\bf q}({\bf {\tilde x}}_{1}),t_{1})
\label{E_Trans1}
\end{equation}
and using Eqs. (\ref{E_Jac}) and (\ref{E_link}) this becomes
\begin{equation}
\rho ({\bf {\tilde x}}_{2}, t_{2} | {\bf {\tilde x}}_{1}, t_{1}) = V |\psi({\bf {\tilde x}}_{2},t_{2})|^{2} \rho_{q} ({\bf q}({\bf {\tilde x}}_{2}),t_{2}|{\bf q}({\bf {\tilde x}}_{1}),t_{1}).
\label{E_Trans2}
\end{equation}
From Eq. (\ref{E_Trans2}) we can see that as the conditional probability density becomes wider and flatter in q-space, $\rho_{q} \rightarrow V^{-1}$, the conditional probability density in physical space approaches QE, $\rho \rightarrow |\psi|^{2}$.\\

This stochastic version of Bohmian mechanics is similar to a stochastic realistic version of quantum mechanics proposed by Nelson \cite{Nelson66} (see also Goldstein \cite{Goldstein87}), but with important differences.
The probability for a given trajectory is not generally the same between the two theories.
However, the most salient difference is that in Nelson's theory, the Brownian motion of the particles cannot cross nodal surfaces, leaving regions isolated by such surfaces out of reach of particle configurations outside of it \cite{Kyprianidis92}.
It is easy to see that this is not the case here since the particles do not `see' the nodal surface in q-space.
They can therefore pass through them and fill any region of physical configuration space, thereby making our theory fully ergodic.
This is not to imply that ergodicity can give a satisfactory account of probabilities in statistical mechanics (see Jaynes \cite{Jaynes67} for a critique).
Nevertheless, in the context of Bohmian mechanics it can help to secure the empirical fact of QE.
This is because although the empirical QE state (in the sense of relative frequencies) occupies most of q-space, should the particles find themselves in a non-QE region they will stay there in Bohmian mechanics whereas in stochastic mechanics they will most probably quickly enter the QE region.
Since relative frequencies inform our beliefs \cite{Cox46}, it follows that the observed empirical QE state will, using Bayes' rule, reinforce the QE hypothesis for the subjective probability density, without the particles actually exploring all of q-space.
In this way, there is a natural compatibility between a stochastic pilot-wave theory and QE for subjective probability without the need for complete ergodicity.\\

Interestingly, the same mathematical analysis found in this section can also be found in Bohm and Vigier (BV) \cite{BohmVigier54}.
They proposed a physical model based on a fluid with irregular random motions to explain QE.
BV considered a particle-like inhomogeneity immersed in a fluid with random fluctuations in the density and stream velocity fields, but whose average is consistent with the Schroedinger equation.
However, BV took their physical model seriously and so refused to consider the possibility that the particle-like inhomogeneity may cross nodal surfaces.
Rather, they denied that such surfaces truly exist.
Whether they exist or not, the fact remains that such crossings are a natural and, to the best of the author's knowledge, unique feature of the density-of-states approach.

\section{Discussion}
\label{S_Disc}

In this work, we have postulated the existence of a q-space representing the set of \emph{a priori} equally likely possible configuration states accessible to an actual configuration of particles.
Quantum mechanics is then seen as a theory about how these possible configuration states relate to the physical configuration space, as a function of time.
The statistical content of quantum theory becomes a consequence of an information theoretic maximum-entropy reasoning.
This not only strengthens the similarities between Bohmian mechanics and the information theoretic interpretaion of statistical mechanics \cite{Jaynes57a,Jaynes57b}, it also allows a Bayesian interpretation of QE in Bohmian mechanics (see Cox \cite{Cox46}, D'Agostini \cite{DAgostini99} and Appleby \cite{Appleby04} for a concise defence of Bayesianism and a critique of frequentism).
This is in constrast to some claims that the minimum Fisher information is better able to describe quantum statistics \cite{Frieden90}.
Rather, it may be that the minimum Fisher information principle only applies to the coordinate transformation itself, allowing the probabilities as such to be described by the maximum-entropy principle (see Reginatto \cite{Reginatto98} and Parwani \cite{Parwani04a,Parwani04b} for more on the use of minimum Fisher information in Quantum mechanics).\\

Indeed, the Jacobian alone does not fully determine the coordinate transformation.
Perhaps there are other principles, not belonging to conventional quantum mechanics, that allow a full determination of the coordinate transformation.
This raises the possibility of a general theory of the mapping of the logically equally possible states to the physical states that contains quantum mechanics as a special case.
Clearly, then, the density-of-states approach raises questions about the origin and nature of this mapping, but it also answers a few questions about Bohmian mechanics.
Since the wavefunction is a by-product of the mapping between the two kinds of configuration states, it stands to reason that it must exist in configurations space, as opposed to ordinary three-dimensional space.
Furthermore, it is easy to see that the relationship between wave and particle in Bohmian mechanics must be asymmetrical, with the wavefunction influencing the particle but not vice-versa (a feature some critics see as a flaw of Bohmian mechanics \cite{Holland93}).
Just as the construction of a die (six or eight-sided, say) has an influence on the actual result obtained upon a throw, but not the other way around, so too does the distribution of equally possible configuration states in the space of physical configuration states influence the actual configuration, but not vice-versa.
It has been argued by D\"urr \emph{et al} \cite{Durr.et.al95} that the dynamical asymmetry and the existence in configuration space is explained if we consider the universal wavefunction as a law in itself and not as a physical object subject to laws.
This view is not incompatible with the density-of-states approach because we can consider every individual trajectory of the mapping, ${\bf x}({\bf q},t)$, as a law.
The states and the laws are therefore linked, and as the particles move in q-space they also move in the space of possible laws.\\

As for the kinematic aspects, the density-of-states approach allows for a range of extensions of the Bohmian guidance condition, subject to a few desirable constraints.
For instance, if one chooses to model motion in q-space as a stochastic process, one may use Einstein or Langevin Brownian motion with the added benefit of complete ergodicity, something not necessarily found in Nelson's model due to nodal surfaces.
It remains to be seen how the density-of-states approach can be implemented for quantum fields, particles with spin and relativistic quantum mechanics.


\begin{thebibliography}{<32>}

\bibitem{Bohm52a} D. Bohm, Phys. Rev. {\bf 85} 166 (1952).
\bibitem{Bohm52b} D. Bohm, Phys. Rev. {\bf 85} 180 (1952).
\bibitem{Holland93} P. R. Holland, {\it The Quantum Theory of Motion} (Cambridge University Press, Cambridge, UK, 1993).
\bibitem{Berndl.et.al95} K. Berndl, M. Daumer, D. D\"urr, S. Goldstein and N. Zanghi, Nuovo Cimento {\bf 110B} 737 (1995).
\bibitem{Shannon48} C. E. Shannon, Bell Sys. Tech. J. {\bf 27} 379 (1948).
\bibitem{Jaynes03} E. T. Jaynes, {\it Probability Theory} (Cambridge University Press, Cambridge, UK, 2003).
\bibitem{Jaynes68} E. T. Jaynes, IEEE Trans. Sys. Sci. and Cybernetics {\bf 4} 227 (1968).
\bibitem{Sklar93} L. Sklar, {\it Physics and Chance} (Cambridge University Press, Cambridge, UK, 1993).
\bibitem{Jaynes63} E. T. Jaynes, in {\it Statistical Physics}, K. W. Ford Ed., (W A Benjamin, Inc., New York, 1963) pp. 163-203.
\bibitem{Jaynes79} E. T. Jaynes, in {\it The Maximum Entropy Formalism}, R. D. Levine and M. Tribus Eds., (M.I.T. Press, Cambridge, MA, 1979) pp. 15-120.
\bibitem{Valentini91a} A. Valentini, Phys. Lett. A {\bf 156} 5 (1991).
\bibitem{Valentini91b} A. Valentini, Phys. Lett. A {\bf 158} 1 (1991).
\bibitem{Valentini04} A. Valentini, e-print hep-th/0407032.
\bibitem{Durr.et.al92} D. D\"urr, S. Goldstein and N. Zanghi, J. Stat. Phys. {\bf 67} 843 (1992).
\bibitem{Lebowitz93} J. L. Lebowitz, Phys. Today {\bf 46} 32 (1993).
\bibitem{Reif65} F. Reif, {\it Fundamentals of Statistical and Thermal Physics} (McGraw-Hill, New York, 1965).
\bibitem{Lamb93} H. Lamb, {\it Hydrodynamics} (Cambridge University Press, Cambridge, UK, 1993).
\bibitem{Nelson66} E. Nelson, Phys. Rev. {\bf 150} 1079 (1966).
\bibitem{Goldstein87} S. Goldstein, J. Stat. Phys. {\bf 47} 645 (1987).
\bibitem{Kyprianidis92} A. Kyprianidis, Found. Phys. {\bf 22} 1449 (1992).
\bibitem{Jaynes67} E. T. Jaynes, in {\it Delaware Seminar in the Foundation of Physics}, M. Bunge Ed., (Springer-Verlag, Berlin, 1967) pp. 77-101.
\bibitem{Cox46} R. T. Cox, Am. J. Phys. {\bf 14} 1 (1946).
\bibitem{BohmVigier54} D. Bohm and J. P. Vigier, Phys. Rev. {\bf 96} 208 (1954).
\bibitem{Jaynes57a} E. T. Jaynes, Phys. Rev. {\bf 106} 620 (1957).
\bibitem{Jaynes57b} E. T. Jaynes, Phys. Rev. {\bf 108} 171 (1957).
\bibitem{DAgostini99} G. D'Agostini, Am. J. Phys. {\bf 67} 1260 (1999).
\bibitem{Appleby04} D. M. Appleby, e-print quant-ph/0408058.
\bibitem{Frieden90} B. R. Frieden, Phys. Rev. A {\bf 41} 4265 (1990).
\bibitem{Reginatto98} M. Reginatto, Phys. Rev. A {\bf 58} 1775 (1998), erratum: Phys. Rev. A {\bf 60} 1730 (1999).
\bibitem{Parwani04a} R. R. Parwani, e-print quant-ph/0408185.
\bibitem{Parwani04b} R. R. Parwani, e-print quant-ph/0412192.
\bibitem{Durr.et.al95} D. D\"urr, S. Goldstein and N. Zanghi, e-print quant-ph/9512031.

\end{thebibliography}
\end{document}